# Stress Evolution in Composite Silicon Electrodes during Lithiation/Delithiation

V.A. Sethuraman,[1] A. Nguyen,[1] M.J. Chon,[1] S.P.V. Nadimpalli,[1] H. Wang,[1]
D.P. Abraham,[2] A.F. Bower,[1] V.B. Shenoy,[1,3] P.R. Guduru[1,*]

[1]School of Engineering, Brown University,
182 Hope Street, Providence, Rhode Island 02912, USA
[2]Chemical Sciences and Engineering Division, Argonne National Laboratory,
Argonne, Illinois 60439, USA
Department of Materials Science and Engineering, University of Pennsylvania,
3231 Walnut Street, Philadelphia, Pennsylvania 19104, USA

[*]Corresponding author, Email: Pradeep_Guduru@Brown.edu, Tel: (1) 401 863 3362

We report real-time average stress measurements on composite silicon electrodes made with two different binders – *viz.* Carboxymethyl cellulose (CMC) and Polyvinylidene fluoride (PVDF) – during electrochemical lithiation and delithiation. During galvanostatic lithiation at very slow rates, the stress in a CMC-based electrode becomes compressive and increases to 70 MPa, where it reaches a plateau and increases slowly thereafter with capacity. The PVDF-based electrode exhibits similar behavior, although with lower peak compressive stress of about 12 MPa. These initial experiments indicate that the stress evolution in a Si composite electrode depends strongly on the mechanical properties of the binder. Stress data obtained from a series of lithiation/delithiation cycles suggests plasticity induced irreversible shape changes in contacting Si particles, and as a result, the stress response of the system during any given lithiation/delithiation cycle depends on the cycling history of the electrode. While these results constitute the first *in situ* stress measurements on composite Si electrodes during electrochemical cycling, the diagnostic technique described herein can be used to assess the mechanical response of a composite electrode made with other active material/binder combinations.

## 1. Introduction

Owing to its high lithiation capacity (*ca.* 3579 mAh/g) and low delithiation potential (*vs.* Li/Li$^+$), silicon is considered to be a promising material to replace graphite as negative electrode in secondary lithium-ion batteries.[1,2] However, Si exhibits large volume expansion and contraction (volumetric strain of 2.7 for Li$_{15}$Si$_4$) during electrochemical lithiation/delithiation cycles; consequently it suffers from mechanical decrepitation during cycling, resulting in rapid capacity fade and poor cycle life. Composite Si electrodes (*i.e.*, Si particles mixed with a polymeric binder and a conductive additive with finite porosity) are well studied in the literature[3-10] and the state-of-the-art Si electrodes that have been shown to cycle well are typically made with a large fraction (typically 80-90%) of micron-sized crystalline Si-based active material (Si powder, carbon-Si mixture, *etc.*) and acetylene black (typically 8%) held together by carboxymethyl cellulose binder.[3] The high Si content and calendaring of the electrodes during fabrication ensure that the Si particles in a composite electrode are in contact with each other, separated by a thin layer of binder and conductive additive.[6,10] Since degradation of such *binder-bridges* results in isolation of particles and capacity-fade, it is necessary to characterize the mechanical interaction between neighboring particles and the stresses in the particles. Although the electrochemical cycling performance of Si composite electrodes has been extensively studied, it appears that stress evolution in Si particles and the average electrode stress during cycling have not been measured thus far. We recently reported *in situ* stress measurements on a silicon thin-film electrode during electrochemical lithiation and delithiation, and showed that the film undergoes cycles of plastic deformation under compressive and tensile stresses upon lithiation and delithiation, respectively.[11-14] When a composite Si electrode is subjected to lithiation and delithiation cycles, it is expected that the contacting regions of neighboring, constrained Si particles will experience compressive stresses and plastic deformation during lithiation and a small tensile stress during delithiation. The magnitude of the tensile stress between particles would be determined and limited by the mechanical properties of the binder and can adversely impact the mechanical integrity of binder-bridges.. Real-time stress measurement during electrochemical cycling is a useful diagnostic tool to quantify the driving force for mechanical degradation of binder bridges and is an essential step towards understanding the mechanics of degradation. Using the substrate-curvature method, we recently reported *in situ* stress measurements on composite graphite electrodes made with MAG-10 graphite particles and PVDF binder, and found that the electrode cycled mostly under a state of compressive stress.[16] Here, we extend this methodology to measure stress evolution during electrochemical lithiation and delithiation of spin-cast composite Si electrodes, in which stresses play a substantially greater role in capacity fade compared to that in graphite electrodes. The spin-casting method results in uniform composite films in the thickness range of 20-40 μm, with good surface coverage and offers flexibility and ease of fabrication compared to traditional battery-electrode casting processes, especially for laboratory testing purposes. Stress response of composite Si electrodes made with two different binders – polyvinylidene fluoride (PVDF), and carboxymethyl cellulose (CMC) – are measured *in situ* during galvanostatic lithiation and delithiation.

## 2. Experimental

### 2.1. Electrode preparation and characterization

Silicon wafers (single-side polished, 50.8 mm diameter, 1.03 mm thick, <111> crystal orientation, with 1.2 μm thermally grown oxide on all sides) were used as substrates for spin-casting composite Si electrodes. The large substrate thickness (~ 1 mm) is necessitated by the need to prevent it from undergoing a buckling instability when the stress-thickness in the composite Si electrode is sufficiently high. In contrast to common application of substrate curvature method where the film thicknesses are usually a fraction of a micron, the active silicon electrode thickness in the present case is about 25 μm. We have shown in our earlier work[11-14] that lithiated silicon can sustain stresses as high as 1 GPa; although the average stress in a particle composite is expected to be substantially lower than 1 GPa, the stress-thickness in such a thick film can potentially cause bifurcation instability in the substrate (for additional information on bifurcation instability, see section 2.6.1. in reference 15). The substrate thickness is chosen to prevent bifurcation according to the criterion discussed in the above reference.

The micron-thick thermal oxide layer was found previously[16] not to alloy with lithium, and hence, isolates the underlying Si wafer substrate from participating in any electrochemical reactions. A thin copper film (*ca.* 200 nm thick) was then sputter deposited (0.013 Pa deposition pressure, 1.5 Å/s deposition rate) onto the unpolished side of the wafer (Lab 18 thin-film deposition system, Kurt J. Lesker Company) to act as the current collector, and also to improve adhesion between the substrate and the composite-electrode layer.[17] Slurry composed of 86 wt. % silicon (325 mesh Si, 99.9% purity, 5 μm average particle size, Sigma Aldrich), 6 wt. % acetylene black (AB) (MTI Corporation), and 8 wt. % polymeric binder, in the form of a pre-made solution, was mixed. PVDF-binder solution was made with 5 wt.% PVDF powder (Sigma Aldrich) dissolved in N-Methyl-2-pyrrolidone (NMP, Sigma Aldrich), while the CMC-binder solution was made with 3.5 wt% sodium CMC powder (Na CMC, Sigma Aldrich) dissolved in de-ionized water. These binder solution ratios were optimized to yield a desirable viscosity for spin coating the slurries. Prior to creating the slurry, the pre-made binder solutions were degassed for 1 hr at a pressure of 0.1 Pa to ensure that the slurry spin coats and evaporates evenly. Prior to coating, the electrode slurry was pulse-sonicated using a Misonix Ultrasonic Liquid Processor S-4000 (Qsonica, Newtown, CT) to create a homogeneous mixture. Pulse-sonication was performed at 50 W power with 30s pulse duration and a duty-ratio of 1 for a total of 10 minutes, which prevents overheating and thickening of the slurry.

The electrode slurry was spin-cast onto the wafer substrates using a spin coater. The CMC slurry was spun cast at 1500 RPM for 45 seconds. Because the PVDF slurry was less viscous, it was spun cast at a lower speed of 1000 RPM for 45 seconds, which resulted in approximately the same coating thickness as that with the CMC binder. Immediately after spin casting, the wafers were placed on a hot plate at 180 °C for approximately 10 minutes to evaporate the solvent. The wafers were then baked in an oven at *ca.* 80°C for 6 hours to remove the solvent from the electrode. The samples were weighed using a microbalance (Mettler Toledo) before and after slurry deposition to determine the amount of silicon deposited on the wafer, which was used in the estimation of the capacity of the silicon electrode. The spin-casting process resulted in a uniform coating with a thickness of *ca.* 30 μm.

The surface morphology and the thickness of the spin-cast electrodes were characterized using a dual-beam scanning electron microscope; the images are shown in Figure 1a and Figure 1c. The porosity of the electrodes is revealed by the focused-ion-beam (FIB) cross-sectional image shown in Figure 1b. The SEM images indicate that the size distribution of the Si particles is between 2-15 μm. In order to visualize how the Si particles that constitute the composite electrode are interconnected to one another by the CMC binder, a thin slice of the electrode (thickness ~ 300 nm) was obtained from a partially lithiated Si/CMC electrode and imaged with a high-resolution transmission electron microscope . An image of the TEM sample is shown in Figure 2a, which reveals the porous nature of the composite electrode. The image also shows several Si particles in contact with their immediate neighbors through binder-bridges such as that shown magnified in Figure 2b.

## 2.2. Electrochemical cell assembly

The electrochemical cell used in this study was made of polytetrafluoroethylene, with a stainless-steel lid and a glass window for substrate-curvature-measurement. A schematic illustration of the cell assembly and a photograph of the experimental setup are shown in Figure 3a-c. Lithium metal (3 mm thick, 52 mm diameter disc, Sigma Aldrich) was placed at the bottom of the cell, and used as both a reference and a counter electrode. Woven Celgard C480 separator (thickness = 21 μm, Celgard Inc., Charlotte, NC) was cut into circular discs measuring 55 mm diameter and placed on top of the metallic Li electrode. The spin-cast Si composite electrode was connected to a thin copper wire (for electrical connectivity) on two diametrically opposite points and placed in the electrochemical cell such that the electrode-side faces the separator while the polished side of the wafer faces up (towards the glass window). The cell was filled with 1.2 M lithium hexafluorophosphate in ethylene carbonate and diethylene carbonate (1:2 by wt.) and was assembled inside a glove-compartment filled with argon. The electrochemical cell used here can be viewed as a flooded beaker cell; and, unlike a coin-cell or a commercial lithium-ion cell, the electrode does not have any confining pressure.

## 2.3. Electrochemical cycling

Electrochemical cycling of the composite Si electrode was carried out in the cell described above in argon atmosphere at 25°C (±1°C) using a Solartron 1470E MultiStat (Solartron Analytical, Oak Ridge, TN). For the first cycle, the composite electrode was galvanostatically lithiated at a current-density of 250 μA/cm$^2$ (ca. C/50 rate) with potential and capacity limits of 0.01 V vs. Li/Li$^+$ and 850 mAh/g, respectively. If the potential limit of 0.01 V was reached before this capacity was achieved, then the current-density was reduced to 125 μA/cm$^2$ to allow for further lithiation of the composite Si electrode. Delithiation was carried out galvanostatically at 125 μA/cm$^2$ until the cell potential reached 1.2 V vs. Li/Li$^+$, followed by a potentiostatic hold at 1.2 V vs. Li/Li$^+$ until the current decayed to less than 1 μA/cm$^2$. Samples have also been subjected to multiple cycles in order to investigate stress evolution under repeated lithiation-delithiation and possible deformation/damage mechanisms. In situ substrate-curvature measurements were performed simultaneously during all lithiation/delithiation cycles.

## 2.4. Real-time stress measurements

In situ stress measurements during electrochemical cycling of composite Si electrodes were performed by monitoring the curvature change of the elastic substrate. The substrate curvature was continuously measured using a multi-beam optical sensor (MOS) wafer-curvature system (k-Space Associates, Dexter, MI).[18,19] The laser source in the MOS system in conjunction with the etalons generates a parallel array of laser beams that is reflected off the sample surface and captured on a CCD camera (Figure 3c). The use of multiple beams alleviates problems associated with system vibrations and improves the signal-noise ratio compared to traditional cantilever beam-deflection methods. The relative change in spot spacing is related to the wafer curvature, $\kappa$, through

$$\kappa = \frac{\left(d - d^0\right)}{d^0}\frac{1}{A_m} \qquad\qquad 1$$

where $d$ is the distance between two adjacent laser spots on the CCD camera [see figure 1(b) in reference 13], $d_0$ is the initial distance and $A_m$ is the mirror constant, given by $2L/cos(\theta)$; $L$ is the optical path length of the laser beam between the plane of the wafer substrate and the CCD array, and $\theta$ is the incident angle of the laser beam on the wafer substrate. The mirror constant $A_m$ is measured by reflecting the laser beam off a flat mirror and a reference mirror of known curvature in the sample plane.

Stoney formula, which relates the stress in a thin film to the curvature of the substrate, is commonly used to convert substrate-curvature data to film stress.[20,21] The Stoney formula is generally valid when the film thickness, $h_f$, is much smaller than that of the substrate $h_s$ (i.e., $h_f \ll h_s$). In the case of composite Si electrodes, the electrode thickness (ca. 32 μm) is not negligible compared to the substrate thickness. Hence, a modified Stoney formula[21] is used to calculate the electrode stress, $\sigma$, from the substrate-curvature $\kappa$

$$\sigma = \frac{M_s h_s^2 \kappa}{6 h_f f(h_i, M_i)} \qquad\qquad 2$$

where $f(h_i, M_i)$ is a function of thicknesses $h_f$, $h_s$ and the biaxial moduli $M_f$ and $M_s$ of the electrode coating and the substrate respectively.

$$f(h_i, M_i) = \left(1 + \frac{h_f}{h_s}\right)\left[1 + 4\frac{h_f}{h_s}\frac{M_f}{M_s} + 6\frac{h_f^2}{h_s^2}\frac{M_f}{M_s} + 4\frac{h_f^3}{h_s^3}\frac{M_f}{M_s} + \frac{h_f^4}{h_s^4}\frac{M_f^2}{M_s^2}\right]^{-1} \qquad\qquad 3$$

The biaxial modulus of the composite electrode was estimated from the elastic constants of its constituents and their volume fraction as

$$M_f = \frac{E_f}{1 - v_f} = \frac{\phi_{Si}E_{Si} + \phi_b E_b + \phi_{AB}E_{AB}}{1 - (\phi_{Si}v_{Si} + \phi_b v_b + \phi_{AB}v_{AB})} \qquad\qquad 4$$

where $\phi_i$, $E_i$ and $v_i$ represent the volume fraction, Young's modulus and Poisson's ratio of component $i$, respectively. Equation 2 reduces to the classical Stoney formula when $f(h_i, M_i) \rightarrow 1$ in the limit as the ratio $h_f / h_s \rightarrow 0$. For the 1040 μm-thick Si wafer substrates used in this study, the sensitivity of the wafer-curvature system is 10 Pa-m. In other words, a 10 μm film with 1 MPa stress can be detected by MOS. In using equation 2 to analyze the experimental data, a constant value of $h_f$ was used, which yields a *nominal* stress, referred to the original thickness. It is a reasonable approximation for a porous electrode for low depths of lithiation. However, for larger depths of lithiation, the evolution of electrode thickness with state-of-charge has to be characterized.

## 3. Results and discussion

The voltage and stress responses of the Si/CMC composite electrode during the first galvanostatic lithiation/delithiation cycle are shown in Figure 4a and 4b, respectively. Upon lithiation, stress increases approximately linearly with state-of-charge to *ca.* 70 MPa at a capacity of *ca.* 600 mAh/g, beyond which it reaches a plateau and continues to increase slowly to a value of *ca.* 77 MPa at *ca.* 850 mAh/g. The region of rapid stress increase is labeled Stage I, and the second region where the stress reaches a plateau is labeled Stage II. Since the electrode is being lithiated for the first time, the lithiation capacities reported include the capacity lost due to formation of the solid-electrolyte-interphase (SEI) layer. The first cycle efficiency suggests that the capacity lost to SEI formation is no more than 78 mAh/g. The cell potential, shown in Figure 4a, is flat (*i.e.*, capacity invariant), which represents the crystalline-Si to amorphous-Li$_x$Si phase boundary propagation on Si particles.[3,14,22,23] The Si/CMC electrode exhibits good reversibility, which is consistent with prior reports,[3,24] and the first-cycle coulombic efficiency (*i.e.,* ratio of delithiation capacity to lithiation capacity) is *ca.* 92%, which is reasonable for a flooded beaker cell. The voltage and stress response of an electrode made with PVDF binder are shown in Figure 5a and Figure 5b respectively. The Si/PVDF electrode exhibited very low first-cycle reversibility (less than 10%) and subsequently, failed to cycle. The potential-capacity data shown in Figure 5a is very similar to data reported by Zhang *et al.* on composite Si electrodes made with PVDF binder.[28] The stress response is qualitatively similar albeit with a lower peak compressive stress of about 12.5 MPa.

In contrast to the Si/CMC electrodes, Si/PVDF electrodes fail prematurely with very low de-lithiation capacity, which is consistent with other reports in the literature.[2,7,25-28] PVDF was shown to bind to Si particles *via* weak van der Waals forces and is known to be ineffective in holding the particles together during large expansion and contraction cycles.[29,30] Moreover, unlike a coin cell, the lack of confining pressure on the electrode in a beaker-cell contributes to particle isolation. It should be noted that the measured stress is the thickness average of the in-plane component of the contact forces between the particles. The local stresses, which depend on the microstructure of the composite, are highly non-uniform as shown by Balke and Kalinin.[31,32] Although detailed mechanics-based modeling is required to interpret the experimental data, some general observations can be made from the data. As the crystalline-amorphous phase boundary propagates into the Si particles, the volume of each particle increases, which results in stresses in regions where particles contact each other. The contact force increases with volume expansion, provided that the binder is strong enough to maintain the integrity of the composite structure. If a composite of idealized regular array of spherical particles is considered, the contact force between a pair of particles is expected to increase non-linearly at the beginning due to elastic deformation and approximately linearly when the particles subsequently undergo elastic-plastic deformation,[33,34] which provides a plausible explanation for the stress evolution observed in Figure 4b in stage I and II. Since the Si particles in the composite electrode are far from being spherical and regular, the inter-particle contact is not symmetric and will necessarily involve shearing and sliding. When the shear strength of the binder is reached, further expansion can be accommodated by inter-particle sliding, which would correspond to the observed plateau in the average stress in stage II of Figure 4b. Since the strength of PVDF is known to be substantially smaller than that of CMC,[42,43] the preceding qualitative explanation is consistent with the observation that the plateau stress for the PVDF-Si composite (Figure 5b) is much less than that in the CMC-Si composite (Figure 4b). It then becomes important to consider the mechanical integrity of binder bridges under shearing and their ability to maintain electrical contact. Recently Bridel *et al.*[35] showed that CMC bonds to Si through a combination of covalent and hydrogen bonds and the former are likely to break during large volume expansion; however, their experiments suggested that the contact between Si and CMC is maintained by hydrogen bonds between –COOH groups of the CMC and the surface SiOH. Bridel *et al.*[35] proposed a self-healing mechanism between CMC and Si that allows large sliding between Si and CMC, which maintains electrical percolation while allowing the particles to deform due to volume expansion. It appears that the plateau stress in Figure 4b represents sliding of binder bridges through the self healing mechanism of Bridel *et al.*[35] and highlights the need for a focused research effort to understand the mechanics of Si-CMC interface.

During lithiation, the Si particles undergo plastic deformation to accommodate volume expansion and consequently, irreversible shape change. As a result, the inter-particle contact force and the average electrode stress are expected to evolve during cycling. In particular, if a lithiation-delithiation cycle is followed by another cycle of smaller depth of lithiation, the electrode stress can be expected to be smaller. Using coupled diffusion-stress finite-element calculations, the plasticity-induced irreversible shape-changes of aggregated Si particles was predicted earlier by Wang *et al.*[36,37]

In the idealized case of a regular array of spherical particles subjected to lithiation/delithiation cycles, the electrode stress during lithiation is expected to remain close to zero until the lithiation capacity equals or exceeds the maximum among the preceding cycles. However, a composite electrode of complex particle geometry is expected to deviate significantly from such an idealized behavior, although the trend is expected to be similar. In order to investigate the influence of plasticity-induced particle shape change on stress evolution, a fresh Si-composite electrode was subjected to a cyclic lithiation-delithiation history shown in Figure 6. The current and potential histories are shown in Figure 6a and 6b respectively and the corresponding stress evolution is shown in Figure 6c. The sample is lithiated to a fixed capacity of *ca.* 240 mAh/g in the first 4 cycles, followed by the 5[th] cycle in which the sample is lithiated to *ca.* 450 mAh/g. The sample is lithiated to the same capacity (*i.e.*, 450 mAh/g) in the 6[th] cycle, followed by an increased capacity of 625 mAh/g in the 7[th] cycle. The electrode experiences small tensile stresses during the initial delithiation half-cycles and this is to the integrity of binder bridges. Note that the tensile stresses disappear when the electrode is subjected to a higher depth of lithiation and delithiation (as seen in Figure 4 as well as in cycle 7 in Figure 6). The potential and stress transients shown in Figure 6 are plotted against capacity in Figure 7.

Focusing attention on cycles 5-7, consider the idealized spherical particle model of the composite in Figure 8a-c. As the particles expand during lithiation, they undergo shape change due to plastic deformation as illustrated in Figure 8a. Figure 8d shows the stress evolution schematically and the lithiated state shown in Figure 8a corresponds to the point A on Figure 8d. If the particles are delithiated from the state shown in Figure 8b, the resulting elastic unloading of the stress is illustrated by the segment AB in Figure 8d. Full delithiation of the particles results in the state shown in Figure 8b and the corresponding stress evolution is represented by the segment BO in Figure 8d. Such a sequence of events represents cycle 5 in Figure 6 and Figure 8e. If the sample is re-lithiated to the previous capacity of point A (Figure 8c), the stress evolution follows the path OBA and subsequent delithiation results in a stress path ABO, which agrees reasonably well with the stress evolution path of cycle 6 in Figure 8e. In the subsequent cycle, as the sample is lithiated to a higher capacity beyond A, the stress evolves along OBAC, which corresponds well with that of cycle 7 in Figure 8e. As noted before, the detailed response is expected to be sensitive to the complex geometry of the particles; however, the general agreement of stress evolution histories suggests the important role played by plasticity and shape changes of particles on stress evolution. In the event of crack formation during delithiation where the composite electrode is biaxially split into several islands (Figure 8c), the stress in each of the islands that are constrained by the substrate contribute to the substrate curvature. This is analogous to early stages of Volmer-Weber growth during thin-film deposition where isolated nuclei under compressive stress contribute to substrate curvature.[38,39] Note that the schematics in Figure 8 should be regarded as highly idealized depiction; the reality is expected to be much more complicated, with a wide particle size distribution and a gradual accumulation of binder-bridge fracture with cycling. The data shown here provides indirect evidence to plasticity-induced shape changes predicted by Wang *et al.*[36,37] A critical issue that arises from the experimental data presented and the preceding discussion is the mechanical integrity of the binder bridges between particles and how the particles maintain electrical contact under multiple lithiation-delithiation cycles; which deserves the attention of the lithium ion battery materials community.

## 4. Conclusions

Direct stress measurements on composite Si electrodes are reported. During the electrochemical lithiation of a Si electrode made with CMC binder, compressive stress increases linearly with state-of-charge reaching *ca.* 70 MPa at *ca.* 600 mAh/g; the stress increases progressively with capacity reaching a value of *ca.* 77 MPa at *ca.* 850 mAh/g. The Si/PVDF electrode exhibits similar behavior although with a lower peak compressive stress of *ca.* 12 MPa. Stress response of a composite Si electrode depends strongly on the mechanical properties of the polymeric binder that constitutes the composite. Stress data obtained from a series of lithiation/delithiation cycles reveal plasticity induced irreversible shape changes in aggregated Si particles, and as a result, the stress response of the system during any given lithiation/delithiation cycle depends on the cycling history of the electrode. The results of the investigation suggest that the focused effort on detailed mechanics of particle deformation and binder-bridges is necessary in order to develop quantitative description of stress evolution and mechanical degradation of silicon-based composite electrodes.

## 5. Acknowledgements

The authors gratefully acknowledge financial support from the United States Department of Energy – EPSCoR Implementation award (grant # DE-SC0007074).

**Table 1: Parameters used for the stress calculations presented in this study.**

| Parameter | Definition | Value | Comments |
|---|---|---|---|
| $d_f$ | Diameter of spin-casted Si electrode | 5.08 cm | Measured |
| $E_{AB}$ | Young's modulus of acetylene black | 10 GPa | Ref. 40 |
| $E_b^{CMC}$ | Young's modulus of Sodium CMC binder | 4 GPa | Ref. 41,42 |
| $E_b^{PVDF}$ | Young's modulus of PVDF binder | 0.65 GPa | Ref. 43 |
| $E_s$ | Young's modulus of Si (111) wafer | 169 GPa | Ref. 47 |
| $E_{Si}$ | Young's modulus of Si particles | 169 GPa | Ref. 44 |
| $f(h_i, M_i)$ | Function in modified Stoney formula | 0.916 | Estimated |
| $h_f$ | Thickness of spin-casted Si electrode | 32 μm | Measured |
| $h_s$ | Thickness of wafer substrate | 1040 μm | Measured |
| $2L/cos(\theta)$ | Mirror constant | 1.9 m | Measured |
| $M_f$ | Biaxial modulus of Si/CMC electrode | 101.37 | Estimated |
| $M_s$ | Biaxial modulus of wafer substrate | 229 | Ref. 47 |
| $\phi$ | Electrode porosity | 0.4 | Assumed |
| $\phi_{AB}$ | Volume fraction of acetylene black | 0.048 | Estimated |
| $\phi_{Si}$ | Volume fraction of Si | 0.516 | Estimated |
| $\phi_b$ | Volume fraction of binder | 0.036 | Estimated |
| $v_{AB}$ | Poisson's ratio of acetylene black | 0.3 | Ref. 45 |
| $v_b^{CMC}$ | Poisson's ratio of Sodium CMC binder | 0.3 | Assumed |
| $v_b^{PVDF}$ | Poisson's ratio of PVDF binder | 0.32 | Ref. 46 |
| $v_{Si}$ | Poisson's ratio of Si particles | 0.21 | Ref. 44 |
| $v_s$ | Poisson's ratio of Si (111) substrate | 0.26 | Ref. 47 |

# Figure-captions

Figure 1: Scanning electron microscopy (SEM) images of as-prepared, spin-casted Si/CMC composite electrodes depicting the (a) surface morphology, (b) steepest wall of a staircase trench dug by a dual-beam focused-ion-beam (FIB) system, and (c) regular cross-section along with the underlying $Si/SiO_2$ wafer substrate. Note that in (b) the angle between the electron-beam and the plane of the trench wall is 52°.

Figure 2: (a) Transmission electron microscopy (TEM) image of a thin slice (*ca.* 300 nm thick) prepared *via* FIB cross-section of a partially lithiated Si/CMC composite electrode and subsequent thinning. (b) Magnified image of the dotted region reveals a pair of neighboring Si particles bridged by the CMC binder.

Figure 3: (a) Layered configuration of the composite Si electrode on Si wafer substrate is shown. Note that this is not drawn to scale. (b) Schematic illustration of the electrochemical cell assembly, and (b) photograph of the apparatus constituting the MOS substrate-curvature-measurement system and the electrochemical-cell assembly.

Figure 4: (a) Potential, and (b) stress response of the spin-casted Si/CMC composite electrode during the first lithiation/delithiation cycle is shown against the capacity. The electrode was lithiated at 250 $\mu A/cm^2$ and 125 $\mu A/cm^2$ (*ca.* C/50 and C/100 rates, respectively) with a lower cut-off potential of 0.01 V *vs.* Li/Li$^+$, and delithiated at 125 $\mu A/cm^2$ with an upper cut-off potential of 1.2 V *vs.* Li/Li$^+$, followed by a potentiostatic delithiation at 1.2 V *vs.* Li/Li$^+$ until the current decayed to less than 1 $\mu A/cm^2$ (*ca.* C/12500 rate).

Figure 5: (a) Potential, and (b) stress response of the spin-casted Si/PVDF composite electrode during the first lithiation/delithiation cycle at 125 $\mu A/cm^2$ (*ca.* C/50 rate) between 0.01 V and 1.2 V *vs.* Li/Li$^+$.

Figure 6: Transient (a) current, (b) potential, and (c) stress response of a Si/CMC composite electrode from an experiment in which the electrode was galvanostatically lithiated at 250 $\mu A/cm^2$ (*ca.* C/50 rate) to a higher state-of-charge during successive sets of cycles followed by galvanostatic delithiation. The duration for each of the galvanostatic lithiation step is indicated above. Delithiation was carried out at 125 $\mu A/cm^2$ for each of the seven cycles until the potential reached 1.2 V *vs.* Li/Li$^+$, followed by a potentiostatic hold at 1.2 V *vs.* Li/Li$^+$ until the current decreased to less than 1 $\mu A/cm^2$ (*ca.* C/12500 rate). Current, potential and stress data corresponding to the first seven lithiation/delithiation cycles are shown. Small tensile stress is present in the initial few cycles due to the integrity of the binder-bridges.

Figure 7: The transient potential and stress data shown in Figure 6 is plotted against the lithiation capacity. For the sake of clarity, only data corresponding to the lithiation process during each of the seven lithiation/delithiation cycles are shown. Excepting the initial lithiation data, the onset for all other lithiation data occurs at a non-zero capacity value because of irreversible loss of capacity in the preceding delithiation half-cycle.

Figure 8: (a)-(c) plasticity-induced irreversible shape changes in contacting silicon particles during the initial lithiation/delithiation cycles is depicted with an idealized spherical particle model. (d) Schematic of stress evolution during a lithiation-delithiation cycle for this idealized case. (e) Stress *vs.* capacity for cycles 5-7 are shown. Both elastic and plastic straining occurs during the lithiation process whereas only elastic strain is recovered during the delithiation process.

**Figure 1**

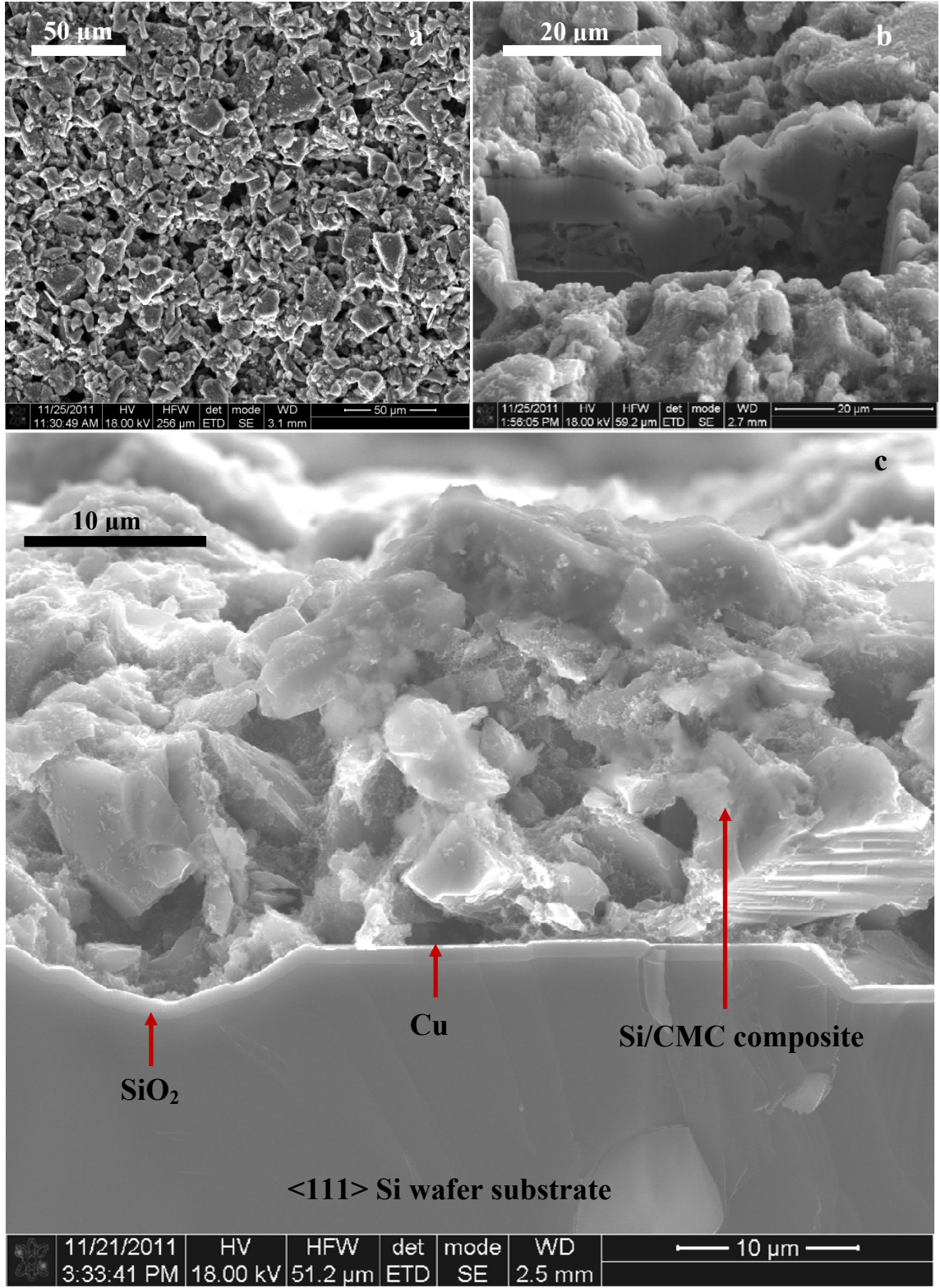

**Figure 2**

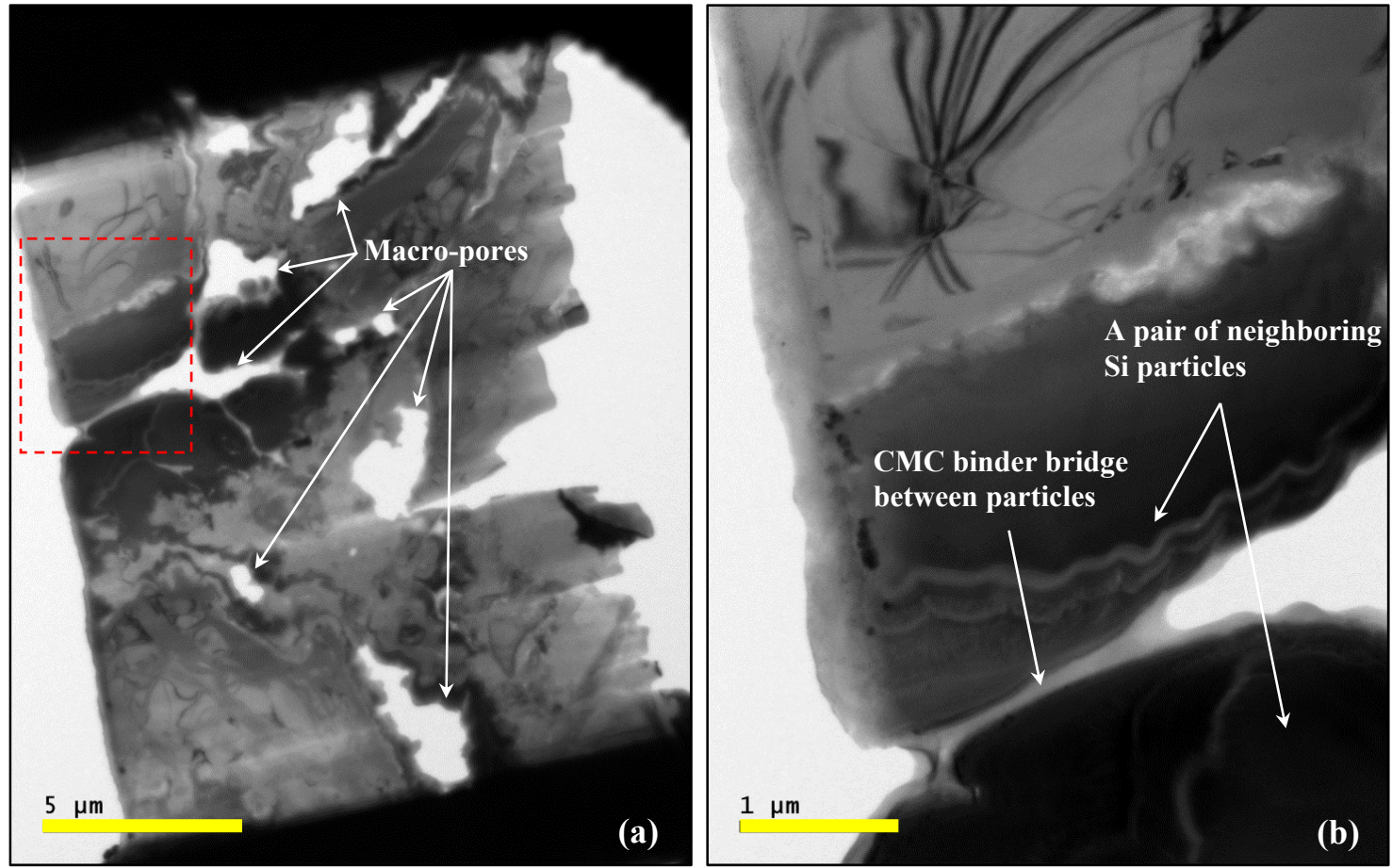

**Figure 3**

**(a)**

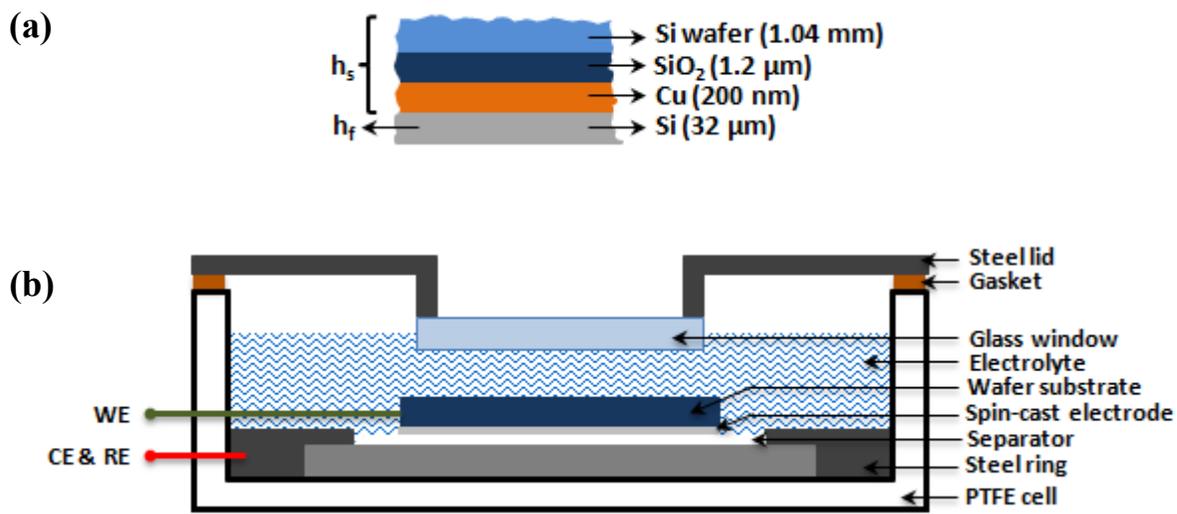

**(b)**

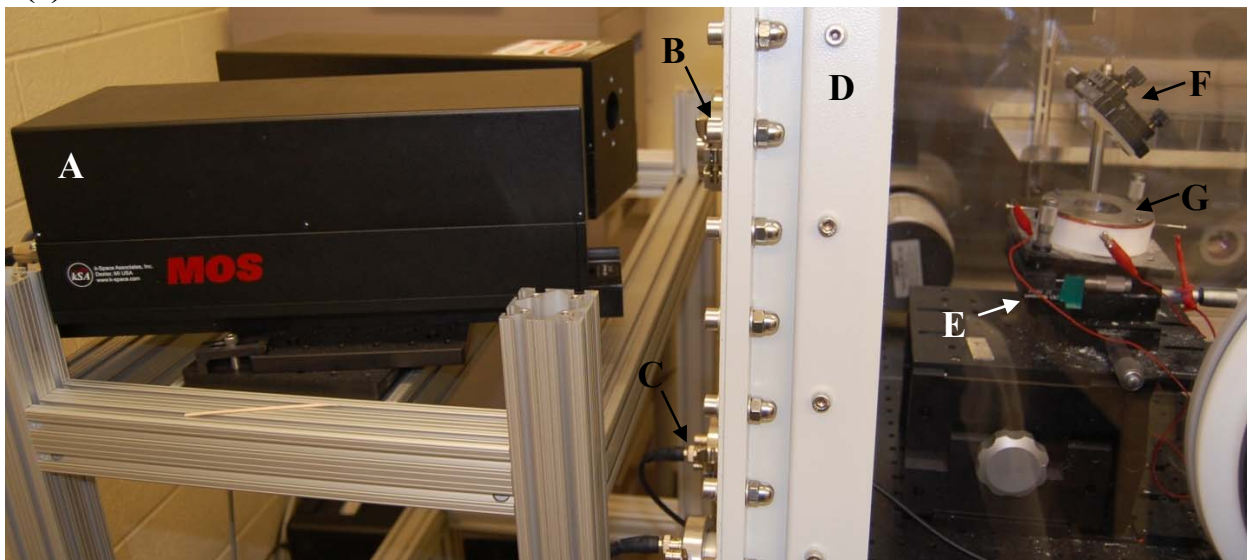

A: Multibeam optical sensor (laser source and CCD camera assembly); B: Optical ports; C: Electrical ports (for communication with the electrochemical cell inside the glove compartment); D: Argon filled glove compartment; E: X-Y tilt stage; F: Alignment mirror; G: Electrochemical cell.

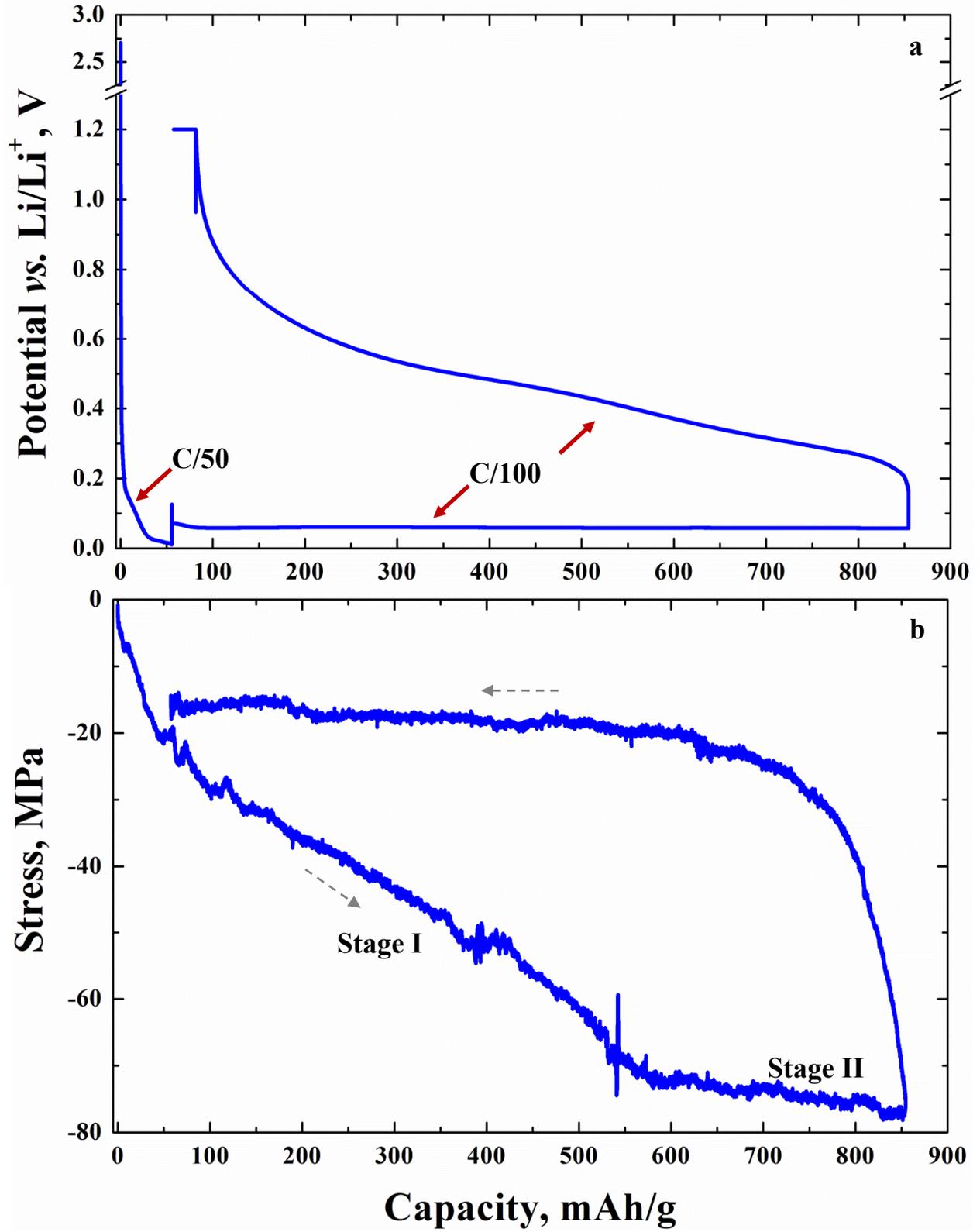

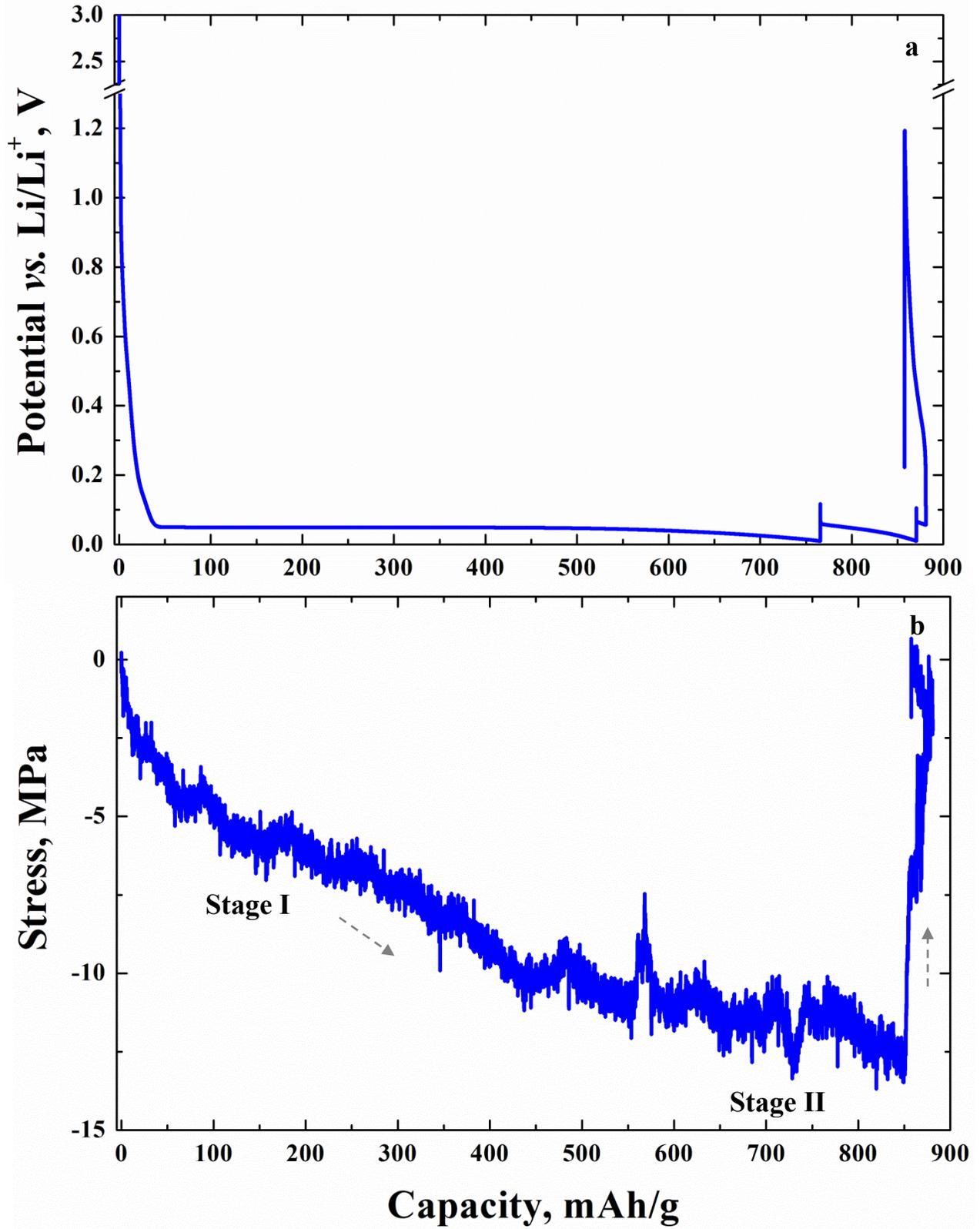

Figure 6

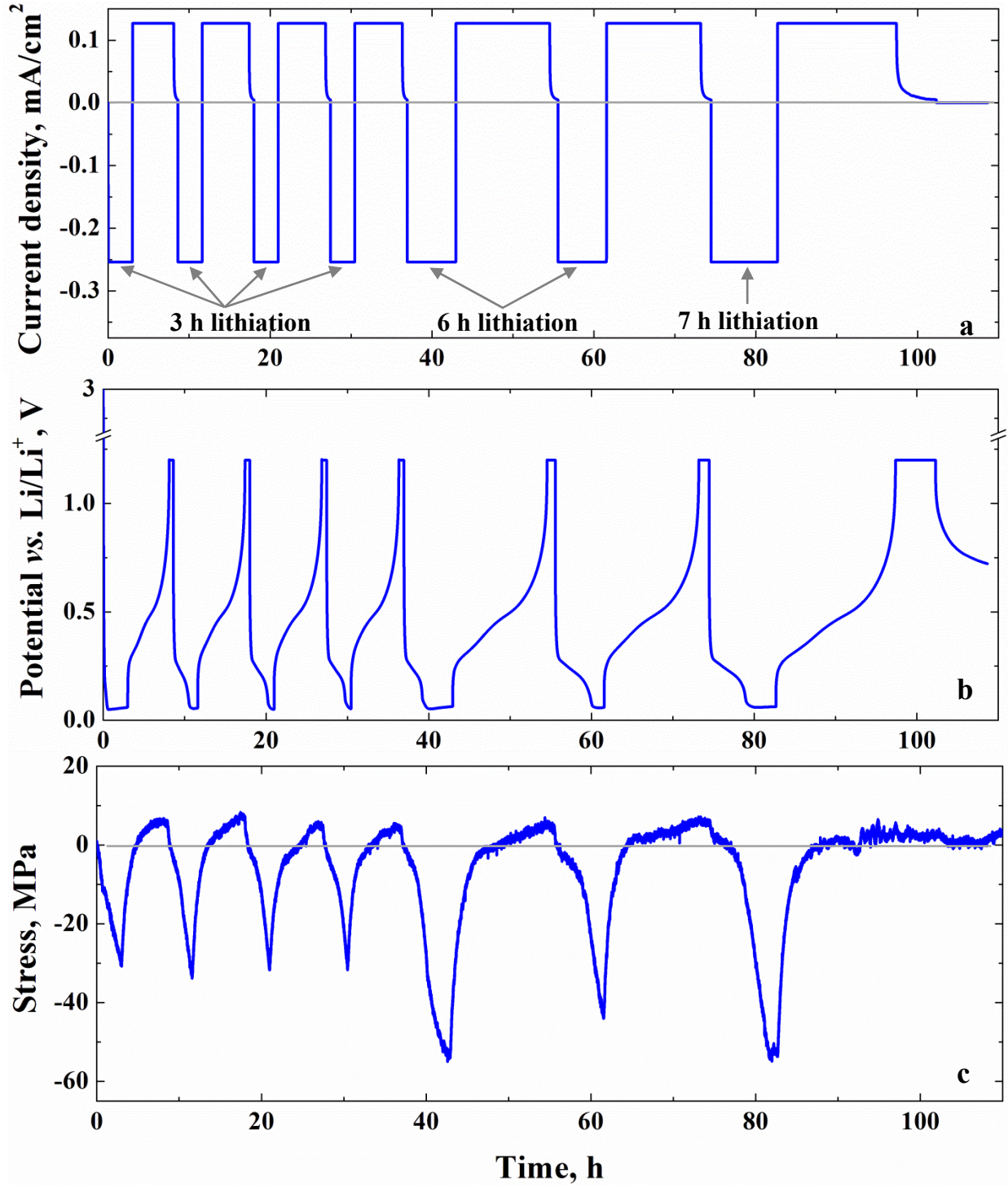

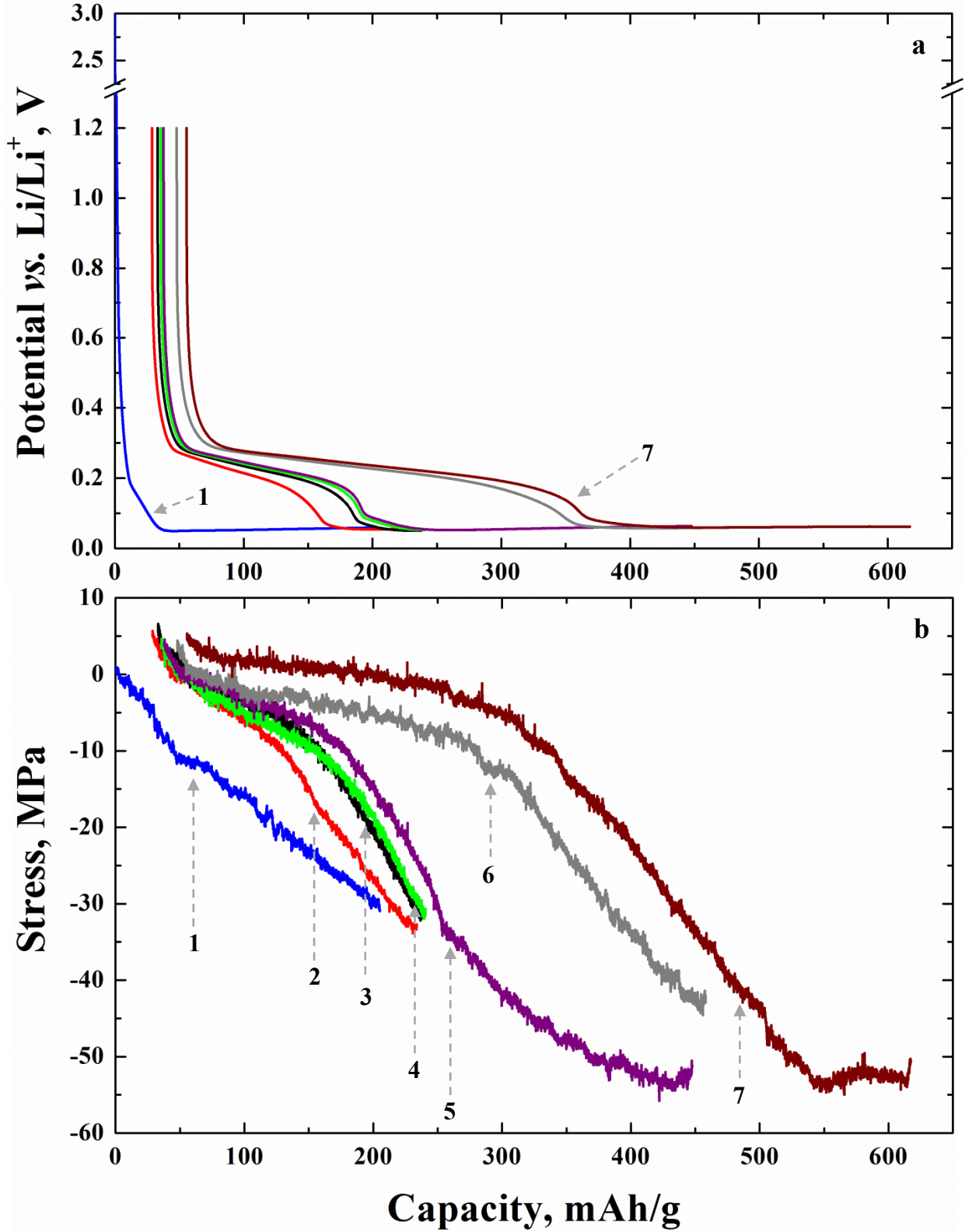

**Figure 8**

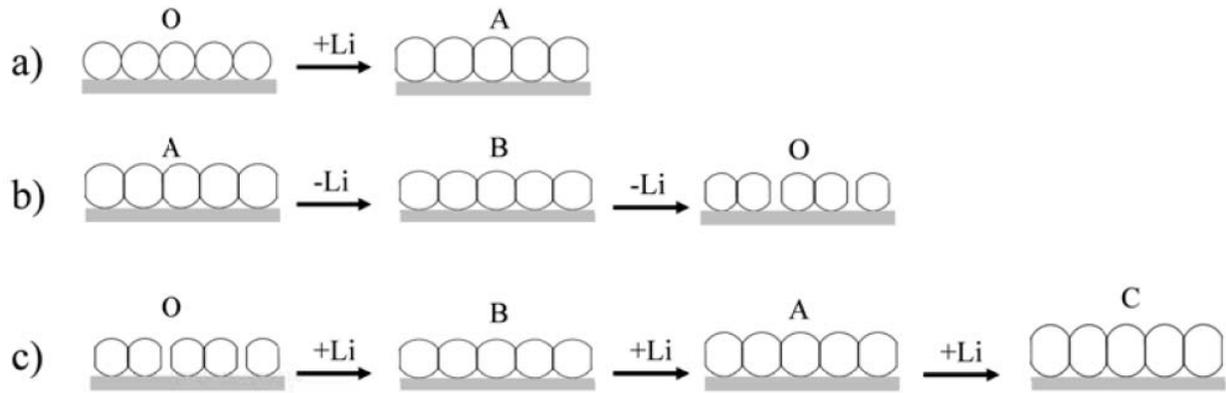

a)

b)

c)

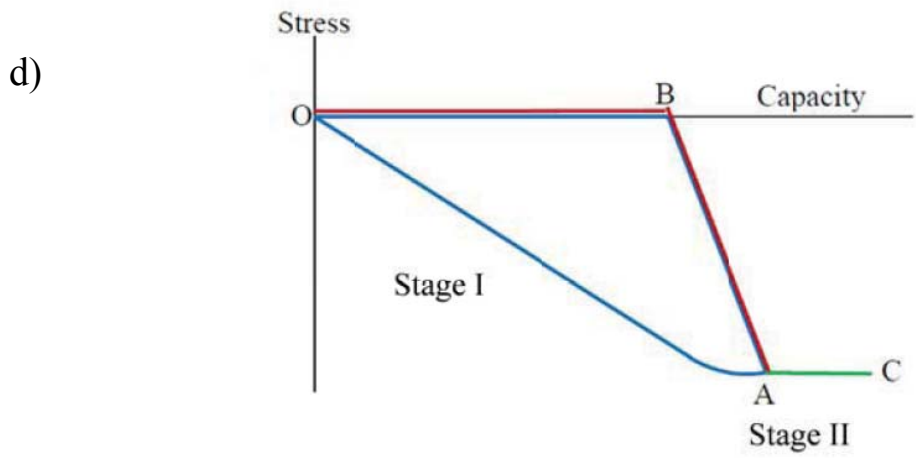

d)

**Figure 8e**

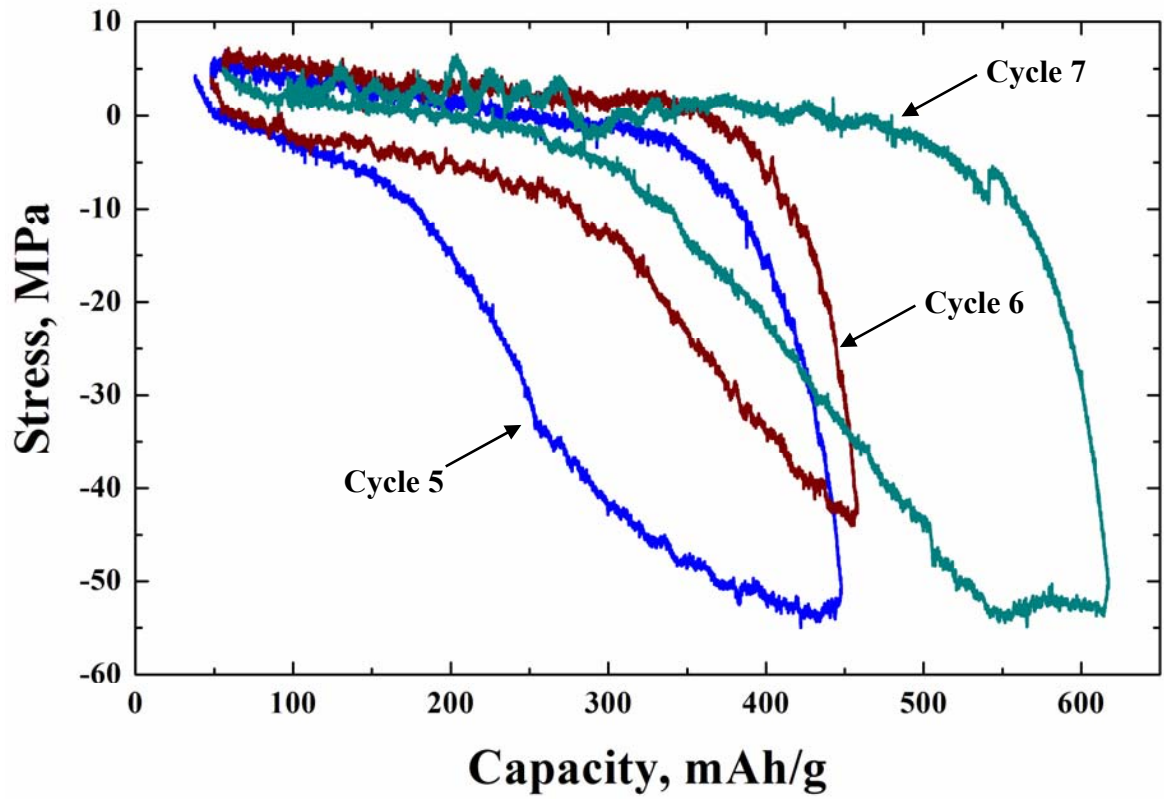